\documentclass[conference]{IEEEtran}
\IEEEoverridecommandlockouts
\usepackage{cite}
\usepackage[dvipdfmx]{graphicx}
\usepackage{amsmath}
\usepackage[dvipdfmx]{color}
\usepackage{bm}
\usepackage{latexsym}
\usepackage{textcomp}
\usepackage[normalem]{ulem}
\usepackage{algorithm}
\usepackage[noend]{algpseudocode}
\usepackage{subfigure}
\usepackage{multirow}
\usepackage{url}
\usepackage{listings}
\usepackage{verbatim}
\usepackage{listings}

\usepackage{xcolor}

\usepackage{subfig}

\def\BibTeX{{\rm B\kern-.05em{\sc i\kern-.025em b}\kern-.08em T\kern-.1667em\lower.7ex\hbox{E}\kern-.125emX}}

\renewcommand{\baselinestretch}{1.0}

\DeclareMathAlphabet{\zch}{OT1}{pzc}{m}{it}

\begin{document}
\title{Cycle-Accurate Evaluation of Software-Hardware Co-Design of Decimal Computation in RISC-V Ecosystem}
\author{\IEEEauthorblockN{Riaz-ul-haque Mian, Michihiro Shintani, and Michiko
   Inoue} \IEEEauthorblockA{ Graduate School of Science and Technology,
   Nara Institute of Science and
    Technology (NAIST)\\ 8916-5 Takayama-cho, Ikoma, Nara, 630-0192 Japan \\
    \{mian.riaz-ul-haque.mn3,shintani,kounoe\}@is.naist.jp} \vspace{-5mm} }

\author{\IEEEauthorblockN{Riaz-ul-haque Mian}
\IEEEauthorblockA{\textit{Graduate School of Information Science} \\
\textit{Nara Institute of Science and Technology}\\
 Ikoma, Japan \\
mian.riaz-ul-haque.mn3@is.naist.jp}
\and
\IEEEauthorblockN{Michihiro Shintani and Michiko Inoue}
\IEEEauthorblockA{\textit{Graduate School of Science and Technology} \\
\textit{Nara Institute of Science and Technology}\\
Ikoma, Japan \\
\{shintani,kounoe\}@is.naist.jp}

}

\maketitle

\newcommand{\pushcode}[1][1]{\hskip\dimexpr#1\algorithmicindent\relax}

\begin{abstract}
 Software-hardware co-design solutions for decimal computation can
 provide several Pareto points to development of embedded systems in
 terms of hardware cost and performance. This paper demonstrates how
 to accurately evaluate such co-design solutions using RISC-V
 ecosystem. In a software-hardware co-design solution, a part of
 solution requires dedicated hardware. In our evaluation framework, we
 develop new decimal oriented instructions supported by an
 accelerator. The framework can realize cycle-accurate analysis for
 performance as well as hardware overhead for co-design solutions for
 decimal computation. The obtained performance result is compared with
 an estimation with dummy functions.
\end{abstract}
\vspace{3mm}

\IEEEpeerreviewmaketitle

\section{Introduction}
 Decimal arithmetic is widely used in financial and
scientific applications. Thus, IEEE 754 (Standard for
floating-point arithmetic) has been revised to include decimal
floating-point formats and operations~\cite{IEEE754}. Many software
(SW) languages support decimal arithmetic that is realized with binary
hardware units. However, these may not be satisfactory for a very
large application in terms of performance.  Many financial
applications need to keep the quality of their customer
service concurrently with the back-end computing process where
computing time is a matter for the business owner.

The decimal arithmetic can be computed with software (arithmetic with
binary hardware units)~\cite{CLib,INTEL,Mar2009tc}, hardware (dedicated hardware unit for decimal
floating-point arithmetic)~\cite{IBMz2011,IBMSCH2001,IBM2009,FUJITSU}, or combination of
both~\cite{Riaz2018}. 
Software solutions are flexible and no additional hardware cost is involved.
Hardware solutions require
high-performance dedicated decimal units with high hardware cost.
Software-hardware co-design solutions can co-optimize flexibility, performance and hardware cost
and give several Pareto points to development of embedded systems. 
In software-hardware co-design solutions, a part of solution requires some dedicated hardware while other part can be executed in standard processors supporting binary arithmetics. 
However, evaluation of co-design solutions requires special evaluation environments.

In~\cite{Riaz2018}, four software-hardware co-design methods for
decimal multiplication are proposed.
part.  A software part is evaluated by running it in several software
platforms by replacing hardware part with dummy functions, while a
hardware part is evaluated by designing hardware with
computer-aided-design tools.  The environment can roughly evaluate the
total performance as an execution time of software program with dummy
functions.

To obtain more accurate evaluation, integrated environment with
dedicated hardware design, software platform, and the interface
between them is required.  An open-source processor like UltraSparc T2
architecture~\cite{OPENSPARC} from Oracle/Sun (the first 64-bit
microprocessors open-sourced) with standard SPARC instruction set
architecture~\cite{SPARCISA} can be used for such evaluation.
However, it requires not only adding new decimal floating-point units
and new instructions for them but also software tools to generate and
simulate binary codes for the new architecture. SPARC V9 architecture
provides IMPDEP1, 2 (Implementation-Dependent Instruction 1,2) and
they can be used for new custom instructions.

In this work, we develop an evaluation framework for software-hardware
co-design of decimal computation using RISC-V ecosystem~\cite{RISCV}.
RISC-V ecosystem is an open-source environment including RISC-V ISA,
Rocket chip (one hardware implementation for RISC-V), RoCC (Rocket
custom co-processor, Rocket chip interface to support accelerators),
several languages for software and hardware, and several tools for
verification and evaluation. In the proposed framework, a co-design
solution is realized as a software that accepts new decimal-oriented
instructions, and the new instructions are supported by a dedicated
accelerator. Cycle-accurate analysis is given by emulating RISC-V
binary on Rocket chip with the dedicated accelerator.

The key contribution of this paper is listed below:

\begin{enumerate}

 \item{Development of an evaluation framework for software-hardware co-design solutions of  decimal computation.}
 \item{Evaluation of software-hardware co-design solution of decimal multiplication.}
 \item{Open-source project of proposed framework available at \texttt{www.decimalarith.info}.}
\end{enumerate}

The organization of the paper is as follows: In
Section~II decimal multiplication using software
hardware co-design are discussed. In section~III the
overview of the proposed framework are presented. The evaluation of
decimal multiplication using the framework is proposed in
Section~IV and the evaluation results are discussed
in~V. Finally, the conclusion is provided in
Section~VI.

\section{Co-Design for Decimal Multiplication}

\begin{figure}[t]
  \centering
  \includegraphics[width=0.75\linewidth]{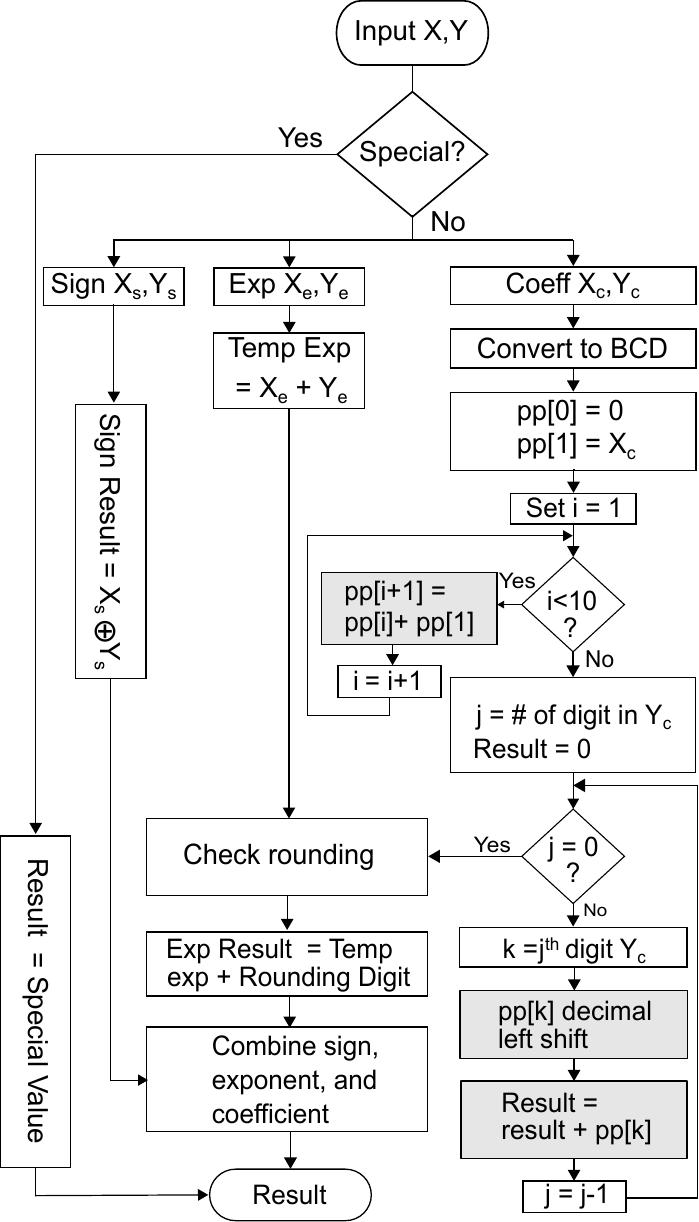}
  \caption{Flow of method-1 in~\cite{Riaz2018}. White and gray blocks
    for software and hardware respectively.}
  \label{fig:method-1}
\end{figure}

Decimal floating-point (DFP) number system represents floating-point
number using base 10. A number is finite or
special value. Every finite number has three integer
parameters sign, coefficient, and exponent.

IEEE 754-2008 compliant decimal multiplication process has the
following basic steps: first, both operands are checked whether they
are a finite number or special values such as NaN (not a number) or
infinity. If either operand is a special value, then the special
general rules are applicable for the operation; otherwise, the
operands are multiplied together with following basic steps:
\begin{itemize}
 \setlength{\itemsep}{-1pt}
 \item {The sign and initial exponent are calculated by doing XOR and
   addition operation between the signs and exponents of multiplier
   and multiplicand.}
\item {Coefficient multiplication is performed to produce the
  coefficient of the result.}
\item {If the result exceeds the precision then rounding operation is
  applied with various rounding algorithm~\cite{Mar2009tc}.}
\item {Finally, the exponents are adjusted accordingly.}
\end{itemize}

The process of such multiplication can be designed using a combination
of software and hardware. Some solutions have been proposed
in~\cite{Riaz2018}.
In this paper, we propose an evaluation framework
  and evaluate Method-1 of~\cite{Riaz2018} as an example.

Figure~\ref{fig:method-1} shows an overall flow of
Method-1 that is one of the solutions in \cite{Riaz2018}.
The method requires one BCD-CLA (BCD
    carry-lookahead adder) as an accelerator to generate multiplicand
  multiples and accumulate partial products.  In addition, no binary
  conversion is required.  To obtain the product of
    coefficients of multiplicand ($X_{c}$) and multiplier ($Y_{c}$),
    these values are first converted into BCD binary-coded
        decimal from DPD (densely packed
        decimal)~\cite{DPD} in software.  The DPD encoding,
   where the coefficient encoding is is very close to BCD
  and can be easily converted to BCD.
  Hardware component BCD adder is used to generate
  multiplicand multiples $1X_{c}$ to $9X_{c}$ by adding
  $X_{c}$ repeatedly. Then the final sum is calculated by adding and
  shifting the multiplicand multiples according to the digit
  of $Y_{c}$. The exponent of the result is finalized by
  adding the number of the rounded digits.

\section{Overview of Framework}
\begin{figure*}[t]
  \centering
  \includegraphics[width=0.82\linewidth]{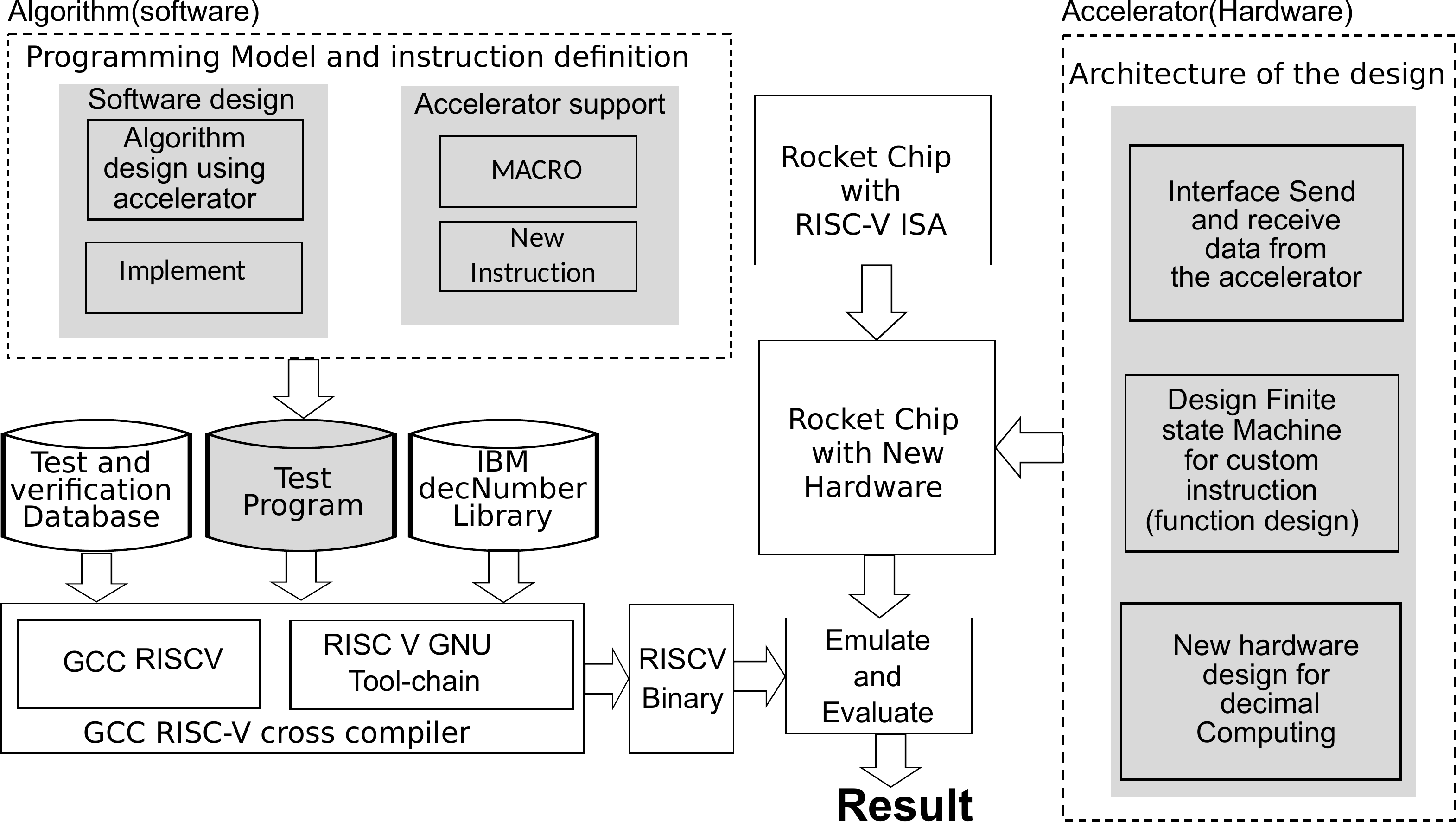}
  \caption{Overview of proposed framework. Gray color indicates our contribution.}
  \label{fig:overall}
\end{figure*}

The proposed framework uses RISC-V ecosystem, decimal C library (decNumber),
arithmetic verification test case and our developed set of test programs. All
major components and software tools, used to integrate the framework,
are listed in Table~\ref{tab:env}. Note that they are fully
open-source programs.
\begin{table}[t]
  \caption{Development environment}
  \label{tab:env}
  \centering
  \scalebox{.95}{
    \begin{tabular}{l|l}
      \hline
       & Description \\ \hline  \hline
      Compiler & GNU GCC for RISC-V as a cross compiler~\cite{GCC}\\ \hline
      ISA simulator & SPIKE~\cite{SPIKE} \\ \hline
      Programming language & High-level: Scale, C++, and C\\
      & Hardware description language: Chisel~\cite{CHISEL} \\
      & Assembly: RISC-V assembly\\ \hline
      ISA & RISC-V~\cite{RISCV} \\ \hline
      Processor core & Rocket~\cite{ROCKET}\\ \hline
      Decimal Software Library & decNumber C library~\cite{CLib}\\\hline
      Testing & Arithmetic operations verification database~\cite{FAHMY2010}\\\hline
  \end{tabular}}
\end{table}

In a co-design methodology, it is very important to decide which functions or operation should have dedicated hardware, and which functions should remain in software to reduce hardware overhead and increase the speed of computing with several tradeoffs. Many parameters including encoding system (BID, DPD), internal format of a decimal number, base, etc. need to be considered for the design.

To design the framework considering software-hardware co-design, we develop a set of hardware components and software program. We include some area efficient hardware components along with associated software that supports decimal computing. 
 Hardware components are realized as a dedicated accelerator.
RISC-V based Rocket core and Rocket custom coprocessor (RoCC) are used in the framework.

The software design may adopt some existing process form~\cite{INTEL,CLib} 
with replacement of some expensive and suitable portion with hardware like~\cite{Riaz2018}, or
a completely new method with new instructions can also be designed. 
In our design, we use base billion, DPD encoding, with BCD-8421 on hardware. However, the design parameter can be flexibly changed by the framework user.

In addition to hardware and corresponding software, we also develop 
a test program generator written in C. The purpose
of the generator is to configure the parameters. 
To use the generator, we first set up some mandatory and optional configurations 
including the format of precision (double or quad), input data-type (rounding, overflow, normal, underflow, etc.), type of the arithmetic operation (addition, subtraction, multiplication or any other), the number of repetition par calculation, pattern of output (execution time or number of cycle) etc.
Then the test program to evaluate target co-design solution is automatically generated.



\section{Evaluation Framework}
\label{sec:propose}

The architecture (hardware) and programming model (software) are described in this section. The overall model of the proposed framework is depicted in Fig.~\ref{fig:overall}. 
On the hardware part, necessary FSM and hardware description for the accelerator are designed. Rocket chip is then compiled and built with the newly generated
accelerator and an executable for the emulator is generated. On the other hand, in software part, RISC-V in-line assembly and C source code are compiled by RISC-V GCC cross compiler to generate RISC-V binary. After that, the binaries are simulated by SPIKE ISA
simulator for functional verification. Hereafter RISC-V machine code
is generated to be executed on the emulator. Finally, 
the emulator is executed to get the evaluation output and wave forms. A detailed description of architecture (hardware) and programming model (software) is given below:

\begin{figure}[t]
	\centering
	\includegraphics[width=0.9\linewidth]{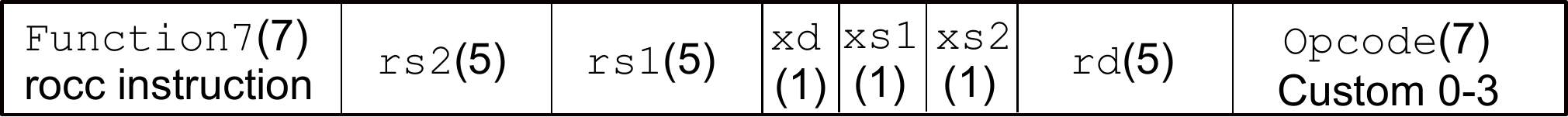}
	\caption{RoCC instruction encoding (number of bits).}
	\label{fig:instruction}
        \vspace{3mm}
\end{figure}

\begin{table}[t]
\centering
\caption{List of instructions}
\label{tlb:function}
\begin{tabular}{l|l|l}\hline
Function    & \textit{Function7}  & Description                                \\ \hline
\texttt{WR} & 0000000 & \begin{tabular}[c]{@{}l@{}}Write a value\\ to a register in Rocket core\end{tabular} \\ \hline
\texttt{RD} & 0000001 & \begin{tabular}[c]{@{}l@{}}Read a value\\ from a register in Rocket core\end{tabular} \\ \hline
\texttt{LD} & 0000010 &  Load a value from a memory\\ \hline
\texttt{ACCUM} & 0000011 &  \begin{tabular}[c]{@{}l@{}}Accumulate a value into a register\\ in Rocket core\end{tabular}\\ \hline
\texttt{DEC\_CNV}   &0000110   & \begin{tabular}[c]{@{}l@{}}Convert binary number to\\ corresponding BCD\end{tabular}\\ \hline
\texttt{DEC\_MUL}   &0000111   & Multiply two BCD numbers               \\ \hline
\texttt{DEC\_ADD}   & 0000100  & Add two BCD numbers                          \\ \hline
\texttt{DEC\_ACCUM} & 0001000  & \begin{tabular}[c]{@{}l@{}}Accumulate BCD numbers stored \\in internal registers\end{tabular}\\ \hline
\end{tabular}
\end{table}

\begin{table*}[t]
\centering
\caption {RoCC instructions (number of bits)}
\label{tlb:instruction}
\begin{tabular}{llllllll}
31                                                                                            & 25                             & 19                            & \multicolumn{3}{l}{15}                                                                                                                                                                                                  & 11                               & 6                                                                                   \\ \hline
\multicolumn{1}{|l|}{Instruction}                                                             & \multicolumn{1}{l|}{Source -1} & \multicolumn{1}{l|}{Source-2} & \multicolumn{3}{l|}{Addressing mode}                                                                                                                                                                                    & \multicolumn{1}{l|}{Destination} & \multicolumn{1}{l|}{Fixed opcode}                                                   \\ \hline
\multicolumn{1}{|l|}{\begin{tabular}[c]{@{}l@{}}\texttt{Function7}(7)\\ rocc instruction\end{tabular}} & \multicolumn{1}{l|}{\texttt{rs2}(5)}    & \multicolumn{1}{l|}{\texttt{rs1}(5)}   & \multicolumn{1}{l|}{\begin{tabular}[c]{@{}l@{}}xd\\ (1)\end{tabular}} & \multicolumn{1}{l|}{\begin{tabular}[c]{@{}l@{}}xs1\\ (1)\end{tabular}} & \multicolumn{1}{l|}{\begin{tabular}[c]{@{}l@{}}xs2\\ (1)\end{tabular}} & \multicolumn{1}{l|}{\texttt{rd}(5)}       & \multicolumn{1}{l|}{\begin{tabular}[c]{@{}l@{}}\texttt{Opcode}(7)\\ Custom-0\end{tabular}} \\ \hline
\multicolumn{1}{|l|}{\texttt{CLR\_ALL} (0000101)}                                                       & \multicolumn{1}{l|}{00000}     & \multicolumn{1}{l|}{00000}    & \multicolumn{1}{l|}{0}                     & \multicolumn{1}{l|}{0}                                                 & \multicolumn{1}{l|}{0}                                                 & \multicolumn{1}{l|}{00000}       & \multicolumn{1}{l|}{0010111}                                                       \\ \hline
\multicolumn{1}{|l|}{\texttt{RD} (0000010)}                                                             & \multicolumn{1}{l|}{00000}     & \multicolumn{1}{l|}{01011}    & \multicolumn{1}{l|}{0}                     & \multicolumn{1}{l|}{0}                                                 & \multicolumn{1}{l|}{1}                                                         & \multicolumn{1}{l|}{00000}       & \multicolumn{1}{l|}{0010111}                                                       \\ \hline
\multicolumn{1}{|l|}{\texttt{WR} (0000000)}                                                             & \multicolumn{1}{l|}{00000}     & \multicolumn{1}{l|}{01011}    & \multicolumn{1}{l|}{1}                      & \multicolumn{1}{l|}{0}                                                 & \multicolumn{1}{l|}{0}                                                        & \multicolumn{0}{l|}{00000}      & \multicolumn{1}{l|}{0010111}                                                       \\ \hline
\multicolumn{1}{|l|}{\texttt{DEC\_ADD} (0000100)}                                                       & \multicolumn{1}{l|}{01010}     & \multicolumn{1}{l|}{01011}    & \multicolumn{1}{l|}{1}                      & \multicolumn{1}{l|}{1}                                                 & \multicolumn{1}{l|}{1}                                                       & \multicolumn{1}{l|}{01100}       & \multicolumn{1}{l|}{0010111}                                                       \\ \hline
\end{tabular}
\vspace{3mm}
\end{table*}

\subsection{Architectural Design (Hardware)}
To embed the decimal arithmetics into a RoCC co-processor, two major parts, interfacing and executing units, are required.
RoCC has three default interface signals and they are:
\begin{enumerate}
\item {Core control (CC): for co-ordination between an accelerator and Rocket core.}
\item {Register mode (Core): for the exchange of data between an accelerator and Rocket core.}
\item {Memory mode (Mem): for communication between an accelerator and L1-D Cache.}
\end{enumerate}
Besides the default interface, there is some extended RoCC interface
like floating-point unit for send and receive data from an FPU and
few more.
The interface between the core and accelerator with default interface signal that communicates through core and memory by the command (\textit{cmd}), response (\textit{resp}).

Figure~\ref{fig:instruction} shows RoCC 32-bit custom instruction encoding with the bit length of several parts of the instruction.
The opcode is selected from {\textit {custom-0} to \textit{custom-3}} reserved for custom instructions, 
and each opcode realizes several functions using \texttt{function7} bits.
Each instruction can have at most one destination register \texttt{rd} and two source registers \texttt{rs1} and \texttt{rs2}.

Three flags \texttt{xd}, \texttt{xs1}, and \texttt{xs2} specify whether registers
in Rocket core are used as destination or source registers. That is,
these flags show the necessity of synchronization between Rocket core
and the accelerator. If the flag value is 1, a register in Rocket core
is used, that is, the values are transferred with the instruction when
{\tt xs1}=``1'' or {\tt xs2}=``1'' and Rocket core waits for the
response when {\tt xd}=``1'', otherwise, it specifies the address of
the register file in the accelerator or the corresponding field is not
used.

Table~\ref{tlb:function} lists the developed  instructions for decimal arithmetic in the framework.
Though we have already developed some of instructions with dedicated hardware, any such hardware component can be integrated into the design.

\begin{figure}[t]
  \centering
  \includegraphics[width=0.95\linewidth]{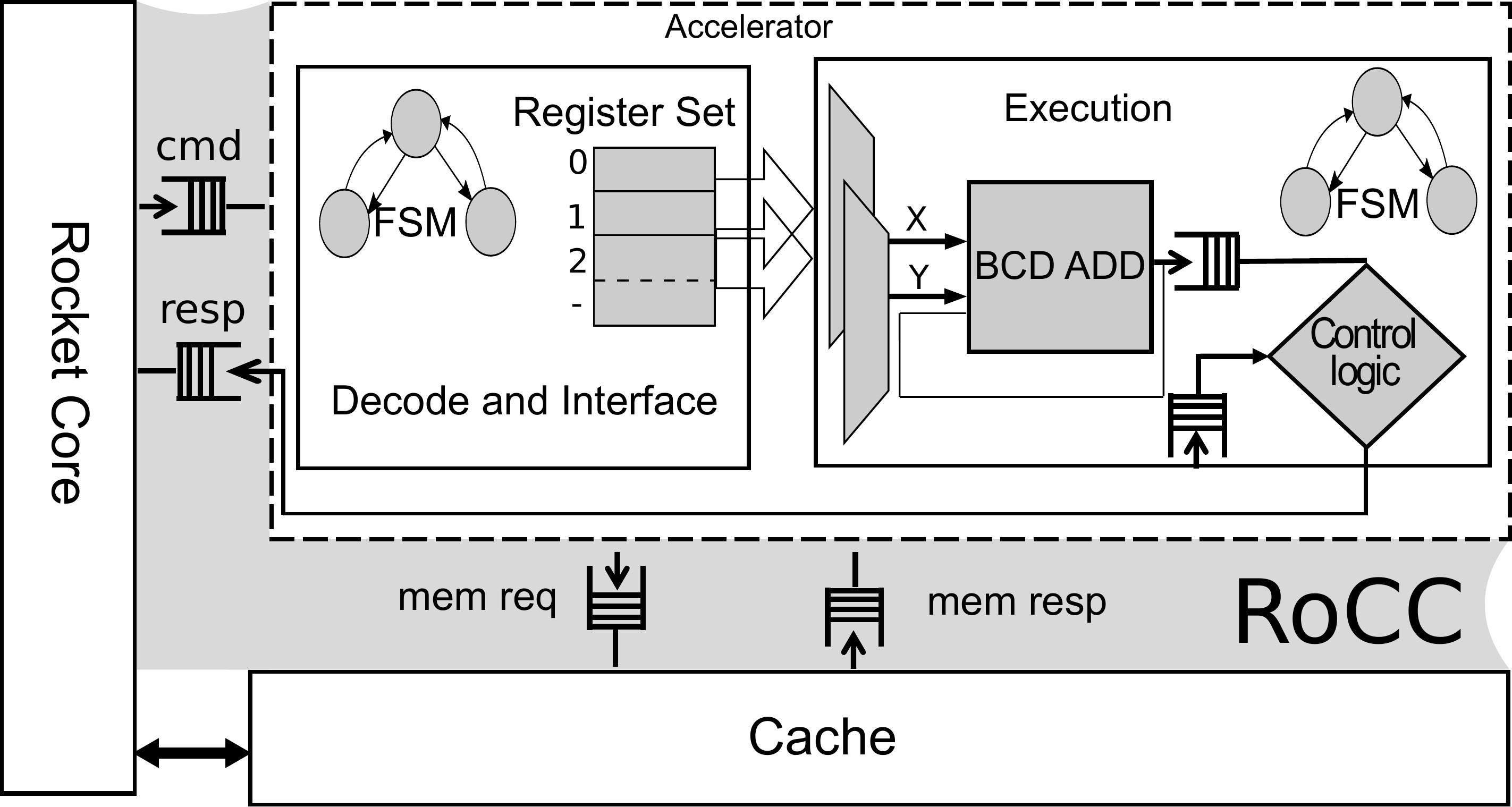}
  \caption{Basic block of Method-1 in~\cite{Riaz2018}.}
  \label{fig:backend}
\end{figure}

\begin{figure}[t]
  \centering
  \includegraphics[width=0.85\linewidth]{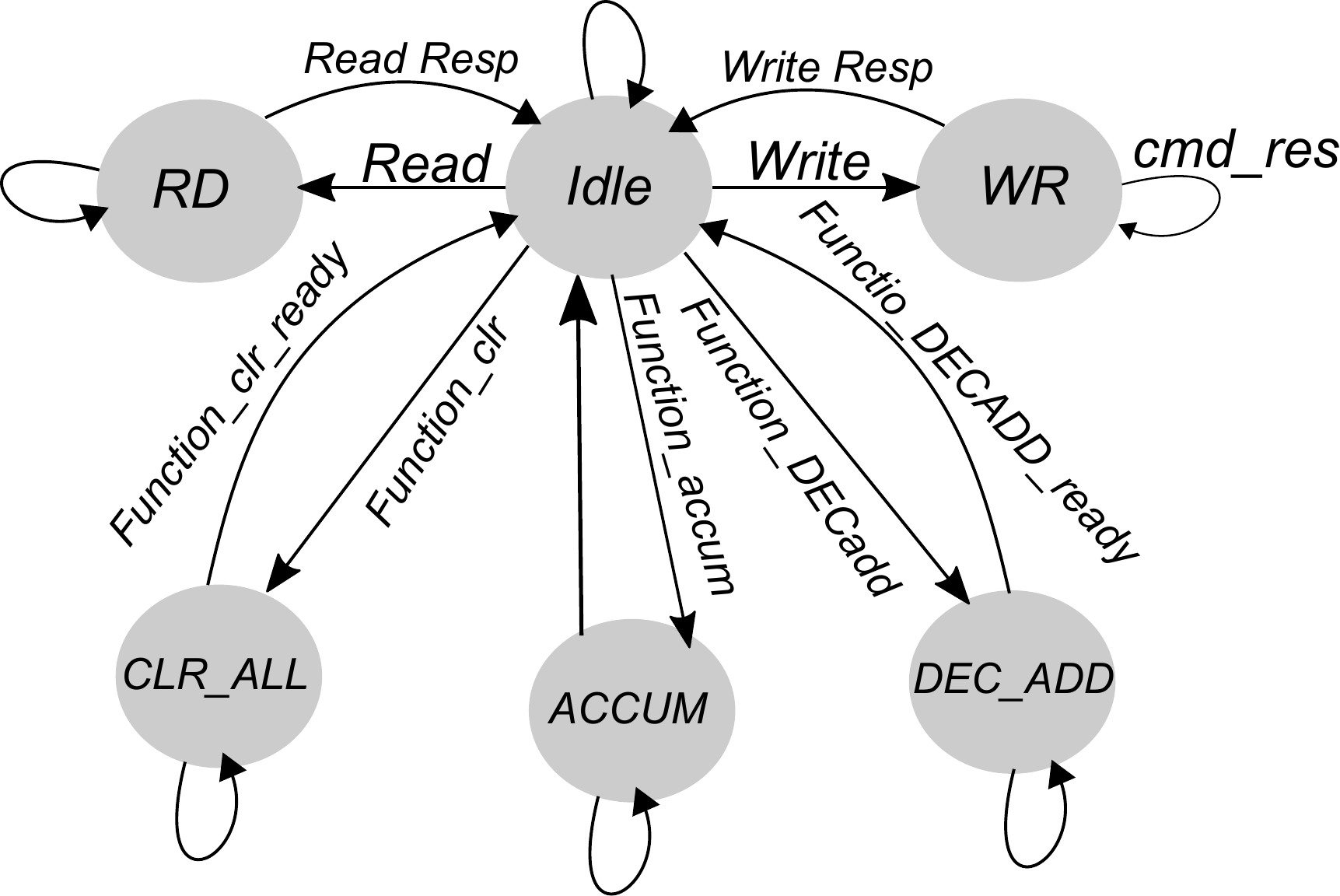}
  \caption{Interface FSM of required function for Method-1 in~\cite{Riaz2018}.}
  \label{fig:fsm}
\end{figure}

The RoCC interface is used to realize communication between Rocket core and the accelerator or between memory (cache) and the accelerator through \textit{cmd} and \textit{resp}. 
The accelerator receives and decodes the commands from Rocket core.
Depending upon the value of \texttt{function7}, 
the corresponding function is executed.
Table~\ref{tlb:instruction} shows the list of some new instructions used in the framework with corresponding values, where core internal integer register 10 and 11 are used as source registers and 12 is used as the destination register. 
Figure~\ref{fig:backend} shows the high-level
architecture of the accelerator, while Fig.~\ref{fig:fsm} shows the
implementation of the finite state machine required to implement the accelerator.

For example,
when (\texttt{DEC\_ADD} instruction is required, it performs BCD
(Binary Coded Decimal) addition for two operands and writes the result
to the destination register.  The pseudo-code written in Chisel for an
instruction \texttt{DEC\_ADD} (\texttt{function7} = ``0000100'') is as
follows.
\begin{lstlisting} %[caption={Define new instruction},captionpos=b][h]
//function bit definition
val DEC_ADD = funct === UInt(4) 
//specific function call
when(cmd.fire() && (DEC_ADD))
//write the result after add X , Y
regfile(addr) := CLA(cmd.bits.rs1,cmd.bits.rs2)
\end{lstlisting}
When the accelerator receives \texttt{DEC\_ADD} command, it executes
the command by adding values of two source registers \texttt{rs1}
and \texttt{rs2} and then write the result to destination
register \texttt{xd}.  Here CLA (Carry-lookahead adder) performs BCD
add operation.  Once the accelerator design is completed, the new chip
is generated with the new accelerator.

\subsection {Programming Model (Software)}
A set of software programs are implemented for the main algorithm where dedicated functions are used to call new instructions supported by the accelerator.

The fragments of the pseudo-code using
accelerator are as follows:

\begin{lstlisting} %[caption={Define new instruction},captionpos=b][h]
//Multiplicand multiple generation 1X-9X
MM[0]=0; MM[1]=X; 
for ( i = 1; i<9 ; i=i+1){
  //Function call in-line assembly
  DEC_ADD_rocc(MM[i+1],MM[1], MM[i]);
}
\end{lstlisting}

\begin{lstlisting} %[caption={Define new instruction},captionpos=b][h]
//Partial porduct accumulation
for ( j=0; j <64 ; j=j+4){
  DEC_ADD_rocc(product,product,MM[MID(Y_64,j,j+4)]);
  product << 4; // shift one decimal digit
}
\end{lstlisting}

The first fragment is used to generate multiplicand multiple and the
second fragment is for accumulating the partial product to generate
final result (See Fig.~\ref{fig:method-1} on Sec.~II).
\textit{DEC\_ADD\_rocc} is a function to call \textit{DEC\_ADD}
instruction.  The function \textit{DEC\_ADD\_rocc} takes two BCD
inputs and return the result after BCD addition.  A set of MACROs and
inline assembly programs are used to define the function
\textit{BCD\_ADD\_rocc}.  In this example code, MID(A, B, C) is MACRO
to extract the specific bit of A within the range C to B. As both
source are in BCD format, where every four bit represents one digit,
thus in every cycle of the for loop pick one digit of multiplier and
then add multiplicand multiple to generate the final product.  The
function that calls \texttt{DEC\_ADD} instruction is as follows.
\begin{lstlisting}
int DEC_ADD_rocc(int a, int b, int c) {
  asm __volatile__ (".word0x08A5F617\n");
  return a;
}
\end{lstlisting}
In the code, "0x08A5F617" is the hex value for
instruction \textit{custom-0} (\texttt{DEC\_ADD}),
where Rocket internal registers 10 and 11 are used as source registers
and 12 is used as the destination register.  In our framework, we also
provide a set of dynamic MACROs to automatically generate the hex
value of corresponding instruction.


\section{Experimental Results} 

\begin{table}[t]
\centering
\caption{Average number of cycles of Method-1 using the framework}
\label{evaluation}
\begin{tabular}{l|r|r|r|r}
\hline
   & SW Part & HW part & Total &Speedup\\ \hline
Method-1~\cite{Riaz2018} & 1013          & 188           & 1201  &2.73x\\ \hline
Software~\cite{CLib} & 3285          & 0             & 3285  &-\\ \hline
\begin{tabular}[c]{@{}l@{}}Method-1 using \\dummy function~\cite{Riaz2018}\end{tabular} & 1446          & 0     &1446  &2.27x \\ \hline
\end{tabular}
\end{table}

\begin{table}[t]
\centering
\caption{Evaluation of Method-1 by real implementation}
\label{separate}
\begin{tabular}{l|r|r}
\hline
 & Time (sec) & Speedup \\ \hline
\begin{tabular}[c]{@{}l@{}}Method-1 using \\dummy function~\cite{Riaz2018}\end{tabular} & 589        & 2.32x \\ \hline
Software~\cite{CLib} & 1367       & -  \\ \hline
\end{tabular}
\end{table}

\begin{table}[t]
\centering
\caption{Evaluation of Method-1 using Gem-5 targeting RISC-V ISA}
\label{tlb:gem5}
\begin{tabular}{l|r|r}
\hline
   & Time (sec) & Speedup \\ \hline
\begin{tabular}[c]{@{}l@{}}Method-1 using \\dummy function~\cite{Riaz2018}\end{tabular}& 0.005443      & 2.30x\\ \hline
Software~\cite{CLib} & 0.012511      & -\\ \hline
\end{tabular}
\end{table}

In~\cite{Riaz2018}, the performance has been evaluated by replacing
the hardware part with a static function called dummy function.  The
dummy functions have a fixed return type and designed according to
methods algorithm. This dummy functions are called from the software
function repeatedly according to method design. This approach is
considered to include interfacing time for software and hardware. Such
an evaluation process has the following limitation:
\begin{enumerate}
  \item Computing time highly dependent on
the nature of the input, like rounding operation takes higher time
than normal operation. However, the dummy function always return a
fixed value and the execution may not follow the expected flow.
  \item The
cycle time for the processor may not the same with and without new
hardware.
\end{enumerate}
Method-1 of~\cite{Riaz2018} with hardware accelerator is implemented
and evaluated total number of cycles with 8,000 sample inputs
including overflow, underflow, normal, rounding, and clamping cases.
For comparison, IBM decNumber C library for double precision decimal
floating-point arithmetic~\cite{CLib} is examined as a software-based
solution, and Method-1 with dummy functions is also examined.  We use
RISC-V RDCYCLE instruction to count the number of
cycles. Table~\ref{evaluation} summarizes the result. From the result,
it is shown that Method-1 with the accerlator is 2.73 times faster
than the software-based solution, while Method-1 with dummy functions
is 2.27 times faster.

The result of Table~\ref{evaluation} is now compared with two other evaluations
to compare a software-based solution (IBM decNumber C library) and Method-1 with dummy functions. 
In Table~\ref{separate}, real implementations of these two methods are executed at Windows 10, 64-bit on Intel core
i7 2.29 GHz with 8 GB RAM. The second evaluation uses Gem-5~\cite{gem5} simulator with AtomicSimpleCPU at system call emulation (SE) mode. In SE mode we need to specify a binary file to be executed of Method-1 of~\cite{Riaz2018}. In this evaluation, we use RISC-V as the target ISA~\cite{RISCV}. Here we also use 8,000 sample and the result is summarizes in Table~\ref{tlb:gem5}. 
From Table~\ref{evaluation} through Table~\ref{tlb:gem5}, it is found that dummy function based evaluations
in three different evaluation exhibit almost the same speedup (2.27x, 2.32x, 2.30x). 
That shows our proposed framework accurately evaluated the cycle times for the target program.
The proposed
framework shows the exact result with 2.73 times improvement in cycle times. The
outcome of the result presents a successful evaluation of the
framework.
Our proposed framework design considering hardware accelerator. Such an
interface imposes a latency overhead during data exchange with CPU
because of the position of the interface into the pipeline. It also
depends on how the core and interface are handled. The impact of such
a problem depends upon the frequency of data exchange between the main
core and the accelerator. Also, due to cache random replacement policy, 
Rocket chip is responsible for computing the number of cycles in
nondeterministically. However, as the framework is proposed for the
integrated evaluation where a large numbers of input samples
with many repetition, the framework can show statistically meaningful results.

\section{Conclusion}
 This paper presents an integrated evaluation framework
using the RISC-V ecosystem, IBM decNumber library, verification test
database with our developed set of test programs. he framework is designed to accurately evaluate software-hardware co-design based decimal computation. A decimal multiplication based on
software hardware co-design is implemented to the framework to validate the
concept of combined decimal multiplication by the actual result. The framework can perform both functional and
behavioral evaluation of any such software-hardware co-design of decimal computation. The framework is an open-source project, and links to all of the source files are available online at \texttt{www.decimalarith.info.}


\bibliographystyle{IEEEtran}
\bibliography{main}

\begin{thebibliography}{10}
\providecommand{\url}[1]{#1}
\csname url@samestyle\endcsname
\providecommand{\newblock}{\relax}
\providecommand{\bibinfo}[2]{#2}
\providecommand{\BIBentrySTDinterwordspacing}{\spaceskip=0pt\relax}
\providecommand{\BIBentryALTinterwordstretchfactor}{4}
\providecommand{\BIBentryALTinterwordspacing}{\spaceskip=\fontdimen2\font plus
\BIBentryALTinterwordstretchfactor\fontdimen3\font minus
  \fontdimen4\font\relax}
\providecommand{\BIBforeignlanguage}[2]{{%
\expandafter\ifx\csname l@#1\endcsname\relax
\typeout{** WARNING: IEEEtran.bst: No hyphenation pattern has been}%
\typeout{** loaded for the language `#1'. Using the pattern for}%
\typeout{** the default language instead.}%
\else
\language=\csname l@#1\endcsname
\fi
#2}}
\providecommand{\BIBdecl}{\relax}
\BIBdecl

\bibitem{IEEE754}
``{IEEE} standard for floating-point arithmetic {IEEE Std} 754-2008,'' pp.
  1--70, 2008.

\bibitem{CLib}
``{C decNumber Library : access date 2018-01-02},'' [Online: \url
  {http://speleotrove.com/decimal/decnumber.html}].

\bibitem{INTEL}
``{IEEE 754-2008 Decimal Floating-Point Compliance Library},'' [Online: \url
  {https://software.intel.com/en-us/articles/intel-decimal-floating-point-math-library}].

\bibitem{Mar2009tc}
M.~Cornea, J.~Harrison, C.~Anderson, and P.~T.~P. Tang, ``A software
  implementation of the {IEEE} {754R} decimal floating-point arithmetic using
  the binary encoding format,'' \emph{IEEE Transactions on Computers}, vol.~58,
  no.~11, pp. 148--162, 2009.

\bibitem{IBMz2011}
S.~Carlough, A.~Collura, S.~Mueller, and M.~Kroener, ``The {IBM}
  z{E}nterprise-196 decimal floating-point accelerator,'' in \emph{Proceedings
  of IEEE International Symposium on Computer Arithmetic}, 2011, pp. 139--146.

\bibitem{IBMSCH2001}
E.~M. Schwarz, J.~S. Kaepernick, and M.~F. Cowlishaw, ``The {IBM} z900 decimal
  arithmetic unit,'' in \emph{Asilomar Conference on Signals Systems and
  Computers}, 2001, pp. 1335--1339.

\bibitem{IBM2009}
E.~M. Schwarz, J.~S. Kapernick, and M.~F. Cowlishaw, ``Decimal floating-point
  support on the {IBM} system z10 processor,'' \emph{IBM Journal of Research
  and Development}, vol.~53, no.~1, pp. 4.1--4.10, 2009.

\bibitem{FUJITSU}
``{Fujitsu's new generation SPARC64 processor},'' [Online: \url
  {http://www.fujitsu.com/global/products/
  computing/servers/unix/sparc-enterprise/key-reports/featurestory/sparce-feature1209.html}].

\bibitem{Riaz2018}
M.~R. ul~haque, M.~Shintani, and M.~Inoue, ``Decimal multiplication using
  combination of software and hardware,'' in \emph{Proceedings of IEEE Asia
  Pacific Conference on Circuits and Systems}, 2018, pp. 239--242.

\bibitem{OPENSPARC}
``{OpenSPARC T1,T2 processor design source code, simulation tools, design
  verification suites, and other tools under open-source licenses},'' [Online:
  \url {http://www.opensparc.net}].

\bibitem{SPARCISA}
``{Oracle SPARC Architecture},'' [Online: \url
  {https://www.oracle.com/technetwork/sparc-architecture-2015-2868130.pdf}].

\bibitem{RISCV}
``{The RISC-V Instruction Set Manual, Volume I: Base User-Level ISA},''
  [Online: \url {https://riscv.org/}].

\bibitem{DPD}
M.~Cowlishaw, ``Densely packed decimal encoding,'' \emph{{IEE Proceedings -
  Computers and Digital Techniques}}, vol. 149, pp. 102--104, 2002.

\bibitem{GCC}
``{GCC, the GNU Compiler Collection},'' [Online: \url {https://gcc.gnu.org/}].

\bibitem{SPIKE}
``{Spike RISC-V ISA Simulator},'' [Online: \url
  {https://github.com/riscv/riscv-isa-sim}].

\bibitem{CHISEL}
J.~Bachrach, H.~Vo, B.~Richards, Y.~Lee, and A.~Waterman, ``{Chisel:}
  constructing hardware in a scala embedded language,'' in \emph{Proceedings of
  IEEE/ACM Design Automation Conference}, 2012, pp. 465--471.

\bibitem{ROCKET}
``{The Rocket Chip Generator},'' [Online: \url {https://riscv.org}].

\bibitem{FAHMY2010}
A.~A.~R. Sayed-Ahmed, H.~A.~H. Fahmy, and M.~Hassan, ``Three engines to solve
  verification constraints of decimal floating-point operations,'' in
  \emph{Asilomar Conference on Signals Systems and Computers}, 2010, pp. 1--4.

\bibitem{gem5}
N.~Binkert, B.~Beckmann, G.~Black, S.~K. Reinhardt, A.~Saidi, A.~Basu,
  J.~Hestness, D.~R. Hower, T.~Krishna, S.~Sardashti, R.~Sen, K.~Sewell,
  M.~Shoaib, N.~Vaish, M.~D. Hill, and D.~A. Wood, ``The {Gem5} simulator,''
  \emph{ACM SIGARCH Computer Architecture News}, vol.~39, no.~2, pp. 1--7,
  2011.

\end{thebibliography}



\end{document}